\documentclass[runningheads]{llncs}

\usepackage{graphicx}
\usepackage{ctable}
\usepackage{multirow}
\usepackage[caption=false]{subfig}

\newcommand{\para}[1]{\vspace{2mm}\noindent\textbf{#1}}

\begin{document}

\title{Popularity Bias in Collaborative Filtering-Based Multimedia Recommender Systems}

\titlerunning{Popularity Bias in CF-Based Multimedia Recommender Systems}

\author{Dominik Kowald \inst{1,2} \and Emanuel Lacic \inst{1}}
\authorrunning{D. Kowald and E. Lacic}

\institute{Know-Center GmbH, Graz, Austria \\\email{{dkowald,elacic}@know-center.at} \and
Graz University of Technology, Graz, Austria}

\maketitle  

\begin{abstract}
Multimedia recommender systems suggest media items, e.g., songs, (digital) books and movies, to users by utilizing concepts of traditional recommender systems such as collaborative filtering. In this paper, we investigate a potential issue of such collaborative-filtering based multimedia recommender systems, namely popularity bias that leads to the underrepresentation of unpopular items in the recommendation lists. Therefore, we study four multimedia datasets, i.e., Last.fm, MovieLens, BookCrossing and MyAnimeList, that we each split into three user groups differing in their inclination to popularity, i.e., LowPop, MedPop and HighPop. Using these user groups, we evaluate four collaborative filtering-based algorithms with respect to popularity bias on the item and the user level. Our findings are three-fold: firstly, we show that users with little interest into popular items tend to have large user profiles and thus, are important data sources for multimedia recommender systems. Secondly, we find that popular items are recommended more frequently than unpopular ones. Thirdly, we find that users with little interest into popular items receive significantly worse recommendations than users with medium or high interest into popularity. 

\end{abstract}

\section{Introduction}
\label{s:intro}
Collaborative filtering (CF) is one of the most traditional but also most powerful concepts for calculating personalized recommendations~\cite{shi_etal:compsurv:2014} and is vastly used in the field of multimedia recommender systems (MMRS)~\cite{deldjoo2020recommender}.   
However, one issue of CF-based approaches is that they are prone to popularity bias, which leads to the overrepresentation of popular items in the recommendation lists~\cite{abdollahpouri2019managing,abdollahpouri2019unfairness}. 
Recent research has studied popularity bias in domains such as music~\cite{kowald2021support,kowald2020unfairness} or movies~\cite{abdollahpouri2019unfairness} by comparing the recommendation performance for different user groups that differ in their inclination to mainstream multimedia items. However, a comprehensive study of investigating popularity bias on the item and user level across several multimedia domains is still missing (see Section~\ref{s:relwork}). 

In the present paper, we therefore build upon these previous works and expand the study of popularity bias to four different domains of MMRS: music (Last.fm), movies (MovieLens), digital books (BookCrossing), and animes (MyAnimeList). 
Within these domains, we show that users with little interest into popular items tend to have large user profiles and thus, are important consumers and data sources for MMRS. 
Furthermore, we apply four different CF-based recommendation algorithms (see Section~\ref{s:method}) on our four datasets that we each split into three user groups that differ in their inclination to popularity (i.e., LowPop, MedPop, and HighPop). With this, we address two research questions (RQ): 

\begin{itemize}
    \item \textbf{RQ1:} To what extent does an item's popularity affect this item's recommendation frequency in MMRS?
    \item \textbf{RQ2:} To what extent does a user's inclination to popular items affect the quality of MMRS?
\end{itemize}

Regarding \textbf{RQ1}, we find that the probability of a multimedia item to be recommended strongly correlates with this items' popularity. 
Regarding \textbf{RQ2}, we find that users with less inclination to popularity (LowPop) receive statistically significantly worse multimedia recommendations than users with medium (MedPop) and high (HighPop) inclination to popular items (see Section~\ref{s:results}). 
Our results demonstrate that although users with little interest into popular items tend to have the largest user profiles, they receive the lowest recommendation accuracy. 
Hence, future research is needed to mitigate popularity bias in MMRS, both on the item and the user level. 

\section{Related Work}
\label{s:relwork}

This section presents research on popularity bias that is related to our work. We split these research outcomes in two groups: (i) work related to recommender systems in general, and (ii) work that focuses on popularity bias mitigation techniques.

\para{Popularity bias in recommender systems.} 
Within the domain of recommender systems, there is an increasing number of works that study the effect of popularity bias. For example, as reported in \cite{baeza2020bias}, bias towards popular items can affect the consumption of items that are not popular. This in turn prevents them to become popular in the future at all. That way, a recommender system is prone to ignoring novel items or the items liked by niche users that are typically hidden in the ``long-tail'' of the available item catalog. 
Tackling these long-tail items has been recognized by some earlier work, such as~\cite{brynjolfsson2006niches,park2008long}.
This issue is further investigated by \cite{abdollahpouri2017controlling,abdollahpouri2019managing} using the popular movie dataset MovieLens 1M. 
The authors show that more than 80\% of all ratings actually belong to popular items, and based on this, focus on improving the trade-off between the ranking accuracy and coverage of long-tail items. Research conducted in~\cite{jannach2015recommenders} illustrates a comprehensive algorithmic comparison with respect to popularity bias. The authors analyze multimedia datasets such as MovieLens, Netflix, Yahoo!Movies and BookCrossing, and find that recommendation methods only consider a small fraction of the available item spectrum. For instance, they find that KNN-based techniques focus mostly on high-rated items and factorization models lean towards recommending popular items. In our work, we analyze an even larger set of multimedia domains and study popularity bias not only on the item but also on the user level.

\para{Popularity bias mitigation techniques.}
Typical research on mitigating popularity bias performs a re-ranking step on a larger set of recommended candidate items. The goal of such post-processing approaches is to better expose long-tail items in the  recommendation list \cite{abdollahpouri2019managing,abdollahpouri2021user,adomavicius2011improving}. Here, for example, \cite{antikacioglu2017post} proposes to improve the total number of distinct recommended items by defining a target distribution of item exposure and minimizing the discrepancy between exposure and recommendation frequency of each item. In order to find a fair ratio between popular and less popular items, \cite{zehlike2017fa} proposes to create a protected group of long-tail items and to ensure that their exposure remains statistically indistinguishable from a given minimum. Beside focusing on post-processing, there are some in-processing attempts in adapting existing recommendation algorithms in a way that the generated recommendations are less biased toward popular items. For example, \cite{adamopoulos2014over} proposes to use a probabilistic neighborhood selection for KNN methods, or \cite{sun2019debiasing} suggests a blind-spot-aware matrix factorization approach that debiases interactions between the recommender system and the user. We believe that the findings of our paper can inform future research on choosing the right mitigation technique for a given setting.

\section{Method}
\label{s:method}
In this section, we describe (i) our definition of popularity, (ii) our four multimedia datasets, and (iii) our four recommendation algorithms based on collaborative filtering as well as our evaluation protocol.

\subsection{Defining Popularity}
Here, we describe how we define popularity (i) on the item level, and (ii) on the user level. We use the item popularity definition of~\cite{abdollahpouri2019unfairness}, where the item popularity score $Pop_i$ of an item $i$ is given by the relative number of users who have rated $i$, i.e., $Pop_i = \frac{|U_i|}{|U|}$. Based on this, we can also define $Pop_{i,u}$ as the average item popularity in the user profile $I_u$, i.e., $Pop_{i,u} = \frac{1}{|I_u|}\sum_{i \in I_u}{Pop_i}$. Additionally, we can also define an item $i$ as popular if it falls within the top-$20\%$ of item popularity scores. Thus, we define $I_{u,Pop}$ as the set of popular items in the user profile.

On the user level, we also follow the work of~\cite{abdollahpouri2019unfairness} and define a user $u$'s inclination to popularity $Pop_u$ as the ratio of popular items in the user profile, i.e., $Pop_u = \frac{|I_{u,Pop}|}{|I_u|}$. As an example, $Pop_u = 0.8$ if 80\% of the items in the user's item history are popular ones.
We use this definition to create the LowPop, MedPop and HighPop user groups in case of MovieLens, BookCrossing and MyAnimeList. In case of Last.fm, we use a definition for $Pop_u$ especially proposed for the music domain, which is termed the mainstreaminess score~\cite{bauer2019global}. Here, we use the $M^{global}_{R,APC}$ definition, which is already provided in the dataset\footnote{\url{https://zenodo.org/record/3475975}} published in our previous work~\cite{kowald2020unfairness}. Formally, $M^{global}_{R,APC}(u) = \tau(ranks(APC),ranks(APC(u)))$, where $APC$ and $APC(u)$ are the artist play counts averaged over all users and for a given user $u$, respectively. $\tau$ indicates the rank-order correlation according to Kendall's $\tau$. Thus, $u$'s mainstreaminess score is defined as the overlap between a user's item history and the aggregated item history of all Last.fm users in the dataset. Thus, the higher the mainstreaminess score, the higher a user's inclination to popular music. Please note that we cannot calculate the mainstreaminess score for the other datasets, since we do not have multiple interactions per item (i.e., play counts) in these cases (only one rating per user-item pair). 

\begin{table}[t]
\setlength{\tabcolsep}{3.5pt}	
\centering
\caption{Statistics of our four datasets, where $|U|$ is the number of users, $|I|$ is the number of media items, $|R|$ is the number of ratings, sparsity is defined as the ratio of observed ratings $|R|$ to possible ratings $|U|\times|I|$, and $R$-range is the rating range.\vspace{-3mm}}
\begin{tabular}{l|rrr|ccc|r}
\specialrule{.2em}{.1em}{.1em}
Dataset & $|U|$ & $|I|$ & $|R|$ & $|R|/|U|$ & $|R|/|I|$ & Sparsity  & $R$-range     \\\hline
    Last.fm      & 3,000 & 352,805 & 1,755,361    & 585       &  5   & 0.998  & [1-1,000]      \\ 
MovieLens      & 3,000 & 3,667 & 675,610      & 225       &  184     & 0.938  & [1-5]         \\
BookCrossing      & 3,000 & 223,607 & 577,414 & 192          & 3     & 0.999      & [1-10]        \\
MyAnimeList    & 3,000 & 9,450 & 649,814      & 216          & 69    & 0.977       & [1-10]        \\
\specialrule{.2em}{.1em}{.1em}
\end{tabular}
\label{tab:stats}
\end{table}

\begin{figure}[t]
\centering
   \subfloat[Last.fm]{
      \includegraphics[width=.45\textwidth]{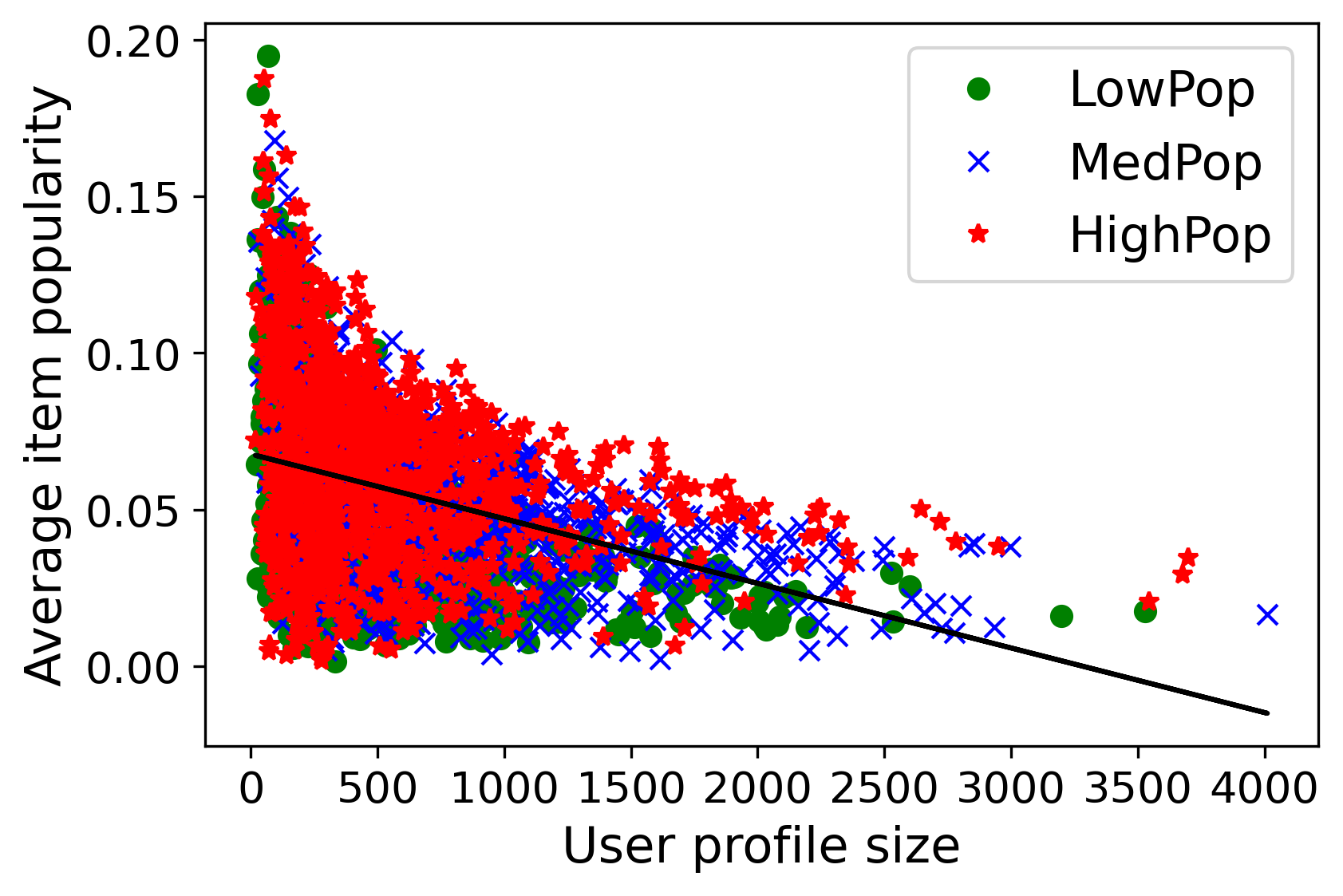}}
~
   \subfloat[MovieLens]{
      \includegraphics[width=.45\textwidth]{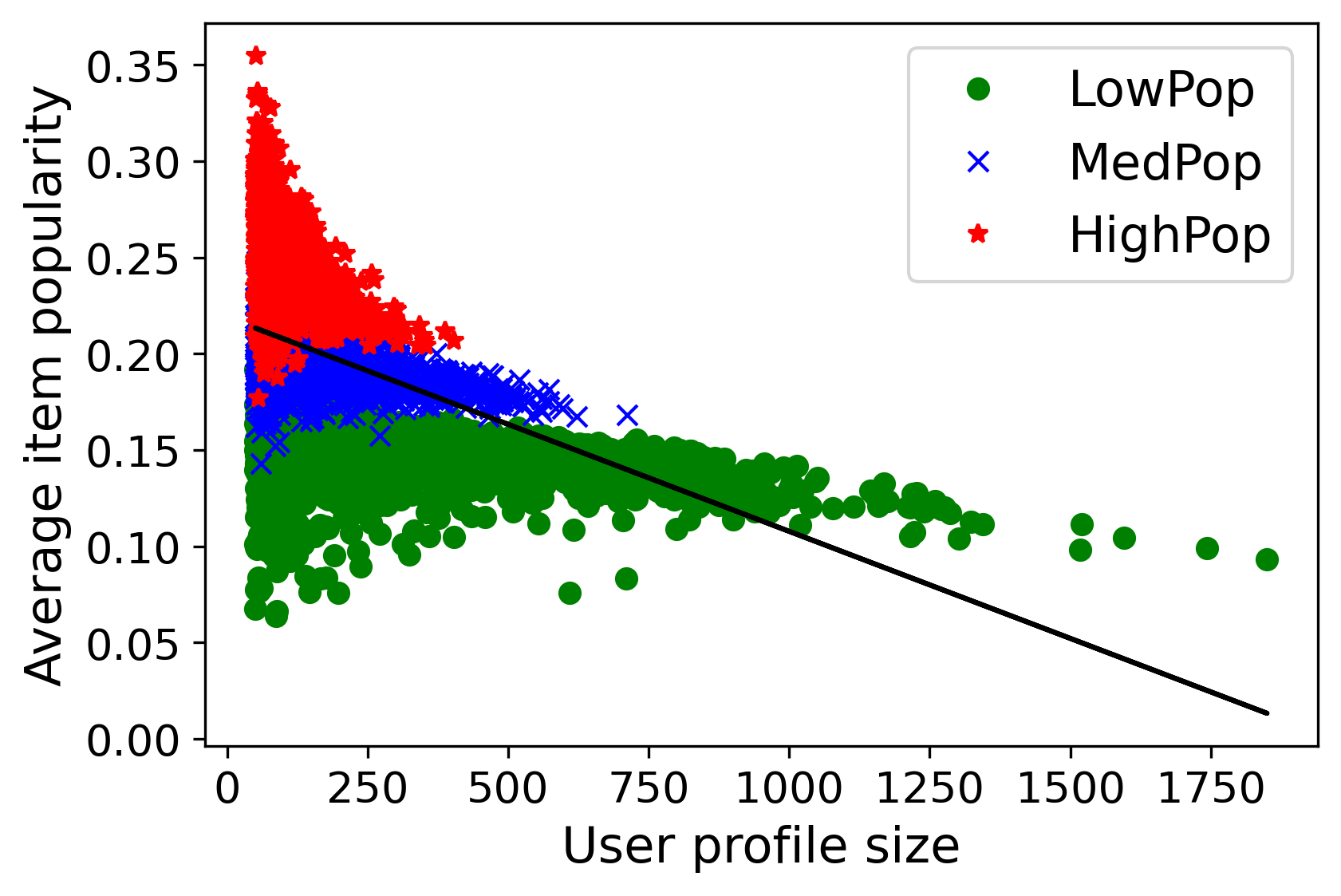}}
\\
   \subfloat[BookCrossing]{
      \includegraphics[width=.45\textwidth]{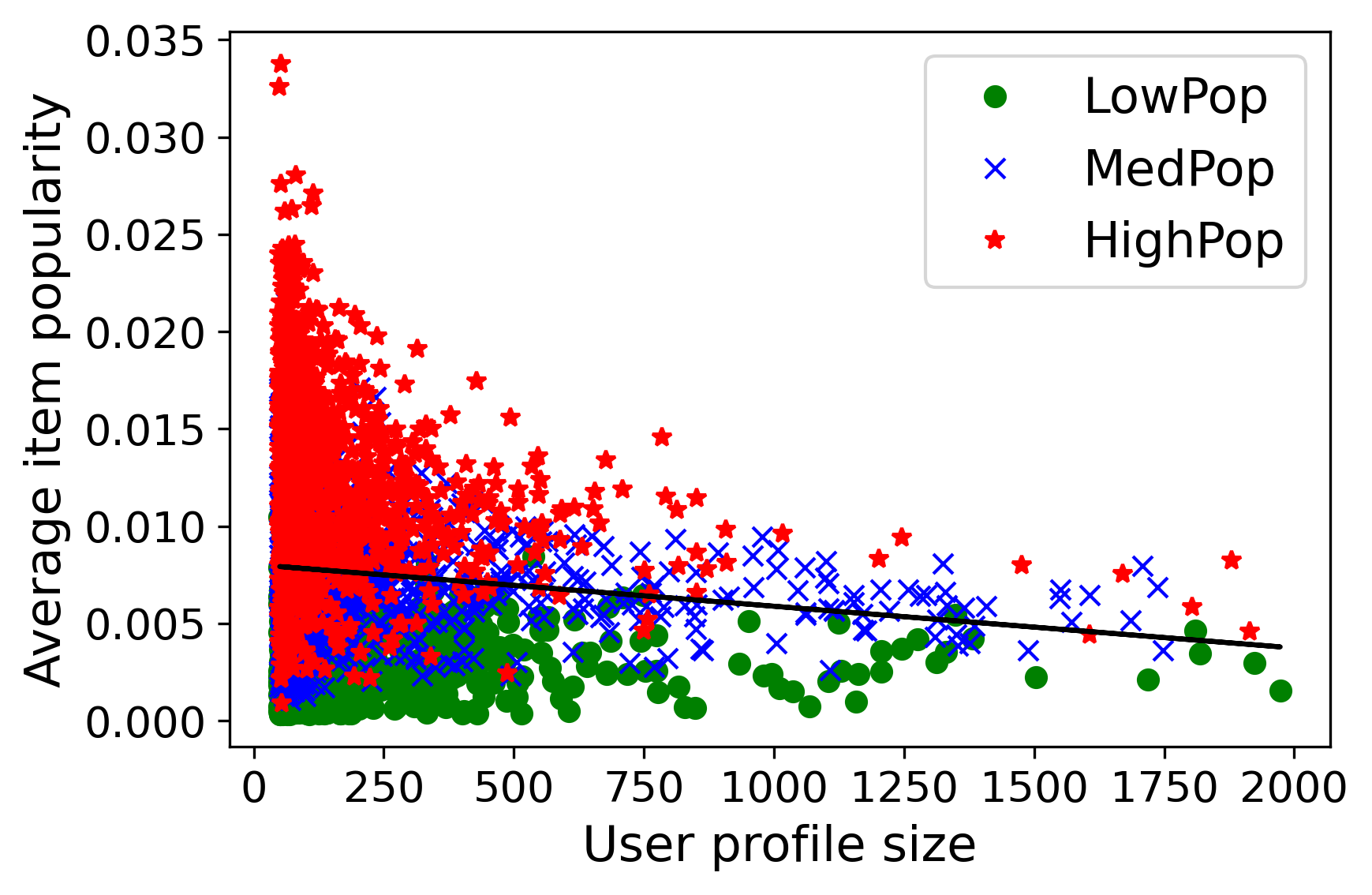}}
~
   \subfloat[MyAnimeList]{
      \includegraphics[width=.45\textwidth]{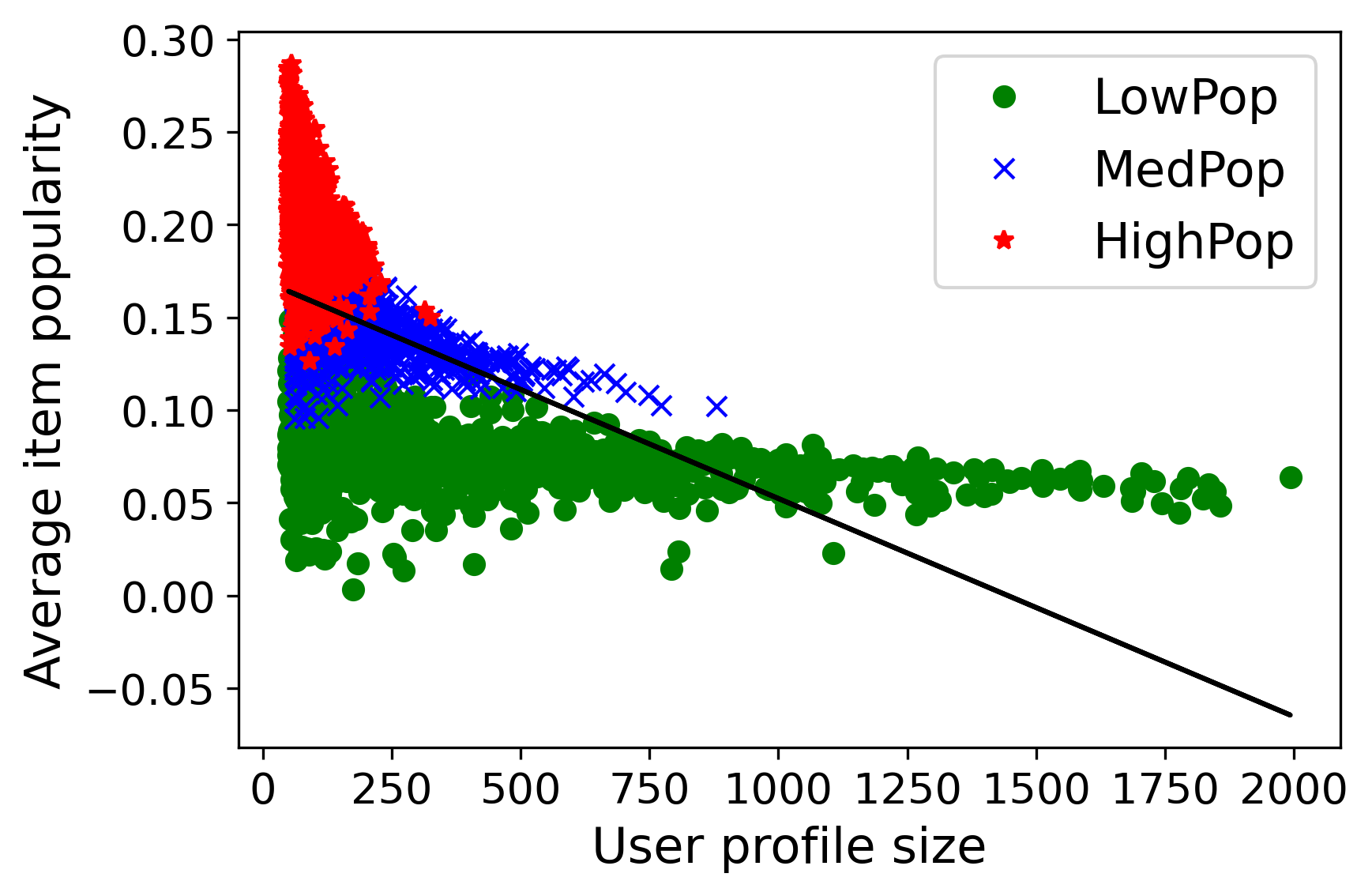}}
   \caption{Relationship between average item popularity scores in the user profiles (i.e., $Pop_{u,i}$) and user profile size (i.e., $|I_u|$). We see that users with little interest in popular items tend to have large user profiles.\vspace{-3mm}}
   \label{fig:pop}
\end{figure}

To get a better feeling of the relationship between average item popularity scores in the user profiles (i.e., $Pop_{u,i}$) and the user profile size (i.e., $|I_u|$), we plot these correlations for our four datasets and per user group in Figure~\ref{fig:pop}. Across all datasets, we see a negative correlation between average item popularity and user profile size, which means that users with little interest in popular items tend to have large user profiles. This suggests that these users are important consumers and data sources in MMRS, and thus, should also be treated in a fair way (i.e., should receive similar accuracy scores as users with medium or high interest in popular items).

\subsection{Multimedia Datasets}
For our study, we use four datasets containing rating data of users for media items. 
The statistics of our datasets can be found in Table~\ref{tab:stats}, and we provide the datasets via Zenodo\footnote{\url{https://zenodo.org/record/6123879}}. 
The users in each of our four datasets are split into three equally-sized user groups: (i) LowPop, i.e., the 1,000 users with the least inclination to popular items, (ii) MedPop, i.e., 1,000 users with medium inclination to popular media items, and (iii) HighPop, i.e., the 1,000 users with the highest inclination to popular media items. This sums up to $|U|=$3,000 users per dataset. Next, we describe our four datasets and how we split the user groups based on the popularity definitions given before:

\para{Last.fm.} For the music streaming platform Last.fm, we use the dataset published in our previous work~\cite{kowald2020unfairness}, which is based on the LFM-1b dataset\footnote{\url{http://www.cp.jku.at/datasets/LFM-1b/}}. Here, a user is assigned to one of the three groups LowPop, MedPop and HighPop based on the user's mainstreaminess score~\cite{bauer2019global}, which we defined earlier (i.e., $M^{global}_{R,APC}$).
Additionally, in this Last.fm dataset, the listening counts of users for music artists are scaled to a rating range of [1-1,000]. 
When looking at Table~\ref{tab:stats}, Last.fm has the largest number of items $|I|=$352,805 and the largest number of ratings $|R|=$1,755,361 across our four datasets.

\para{MovieLens.} In case of the movie rating portal MovieLens, we use the well-known MovieLens-1M dataset\footnote{\url{https://grouplens.org/datasets/movielens/1m/}}. We extract all users with a minimum of 50 ratings and a maximum of 2,000 ratings. We assign these users to one of the three user groups LowPop, MedPop and HighPop based on the ratio of popular items in the user profiles~\cite{abdollahpouri2019unfairness} as described earlier (i.e., $Pop_u$). 
Table~\ref{tab:stats} shows that MovieLens is the least sparse (i.e., most dense) dataset in our study and also has the highest number of ratings per items ($|R|/|I|$).

\para{BookCrossing.} The dataset of the (digital) book sharing platform BookCrossing was provided by Uni Freiburg\footnote{\url{http://www2.informatik.uni-freiburg.de/~cziegler/BX/}}. We use the same popularity definitions, group assignment method as well as rating thresholds as in case of MovieLens. However, in contrast to MovieLens, BookCrossing contains not only explicit feedback in the form of ratings but also implicit feedback when a user bookmarks a book. In this case, we set the implicit feedback to a rating of 5, which is the middle value in BookCrossing's rating range of [1-10]. Across all datasets, BookCrossing is the dataset with the highest sparsity. 

\para{MyAnimeList.} 
We apply the same processing methods as used in case of BookCrossing to the MyAnimeList dataset, which is provided via Kaggle\footnote{\url{https://www.kaggle.com/CooperUnion/anime-recommendations-database}}. 
Similar to BookCrossing, MyAnimeList also contains implicit feedback when a user bookmarks an Anime, and again we convert this feedback to an explicit rating of 5, which is the middle value in the rating range.

\subsection{Recommendation Algorithms and Evaluation Protocol}
We use the same set of personalized recommendation algorithms as used in our previous work~\cite{kowald2020unfairness} but since we focus on CF-based methods, we replace the UserItemAvg algorithm with a scalable  co-clustering-based approach~\cite{george2005scalable} provided by the Python-based Surprise framework\footnote{\url{http://surpriselib.com/}}. Thus, we evaluate two KNN-based algorithms without and with incorporating the average rating of the target user and item (UserKNN and UserKNNAvg), one non-negative matrix factorization variant~\cite{luo2014efficient} (NMF) as well as the aforementioned CoClustering algorithm. In most cases, we stick to the default parameter settings as suggested by the Surprise framework and provide the detailed settings in our GitHub repository\footnote{\url{https://github.com/domkowald/FairRecSys}}.

We also follow the same evaluation protocol as used in our previous work~\cite{kowald2020unfairness} and formulate the recommendation task as a rating prediction problem, which we measure using the mean absolute error (MAE). However, instead of using only one 80/20 train-set split, we use a more sophisticated 5-fold cross-validation evaluation protocol. To test for statistical significance, we perform pairwise t-tests between LowPop and MedPop as well as between LowPop and HighPop since we are interested if LowPop is treated in an unfair way by the MMRS. We report statistical significance for LowPop only in cases in which there is a significant difference between LowPop and MedPop as well as between LowPop and HighPop for all five folds.

\section{Results}
\label{s:results}

We structure our results based on our two research questions. Thus, we first investigate popularity bias on the item level by investigating the relationship between item popularity and recommendation frequency (\textbf{RQ1}). Next, we investigate popularity bias on the user level by comparing the recommendation performance for our three user groups (\textbf{RQ2}).

\begin{figure}[t]

\begin{center}
 \scriptsize
\begin{tabular}{c c c c c}
   \multicolumn{1}{c}{UserKNN}    & \multicolumn{1}{c}{UserKNNAvg} & \multicolumn{1}{c}{NMF}  & \multicolumn{1}{c}{CoClustering}  \\ 
\includegraphics[width=.24\textwidth]{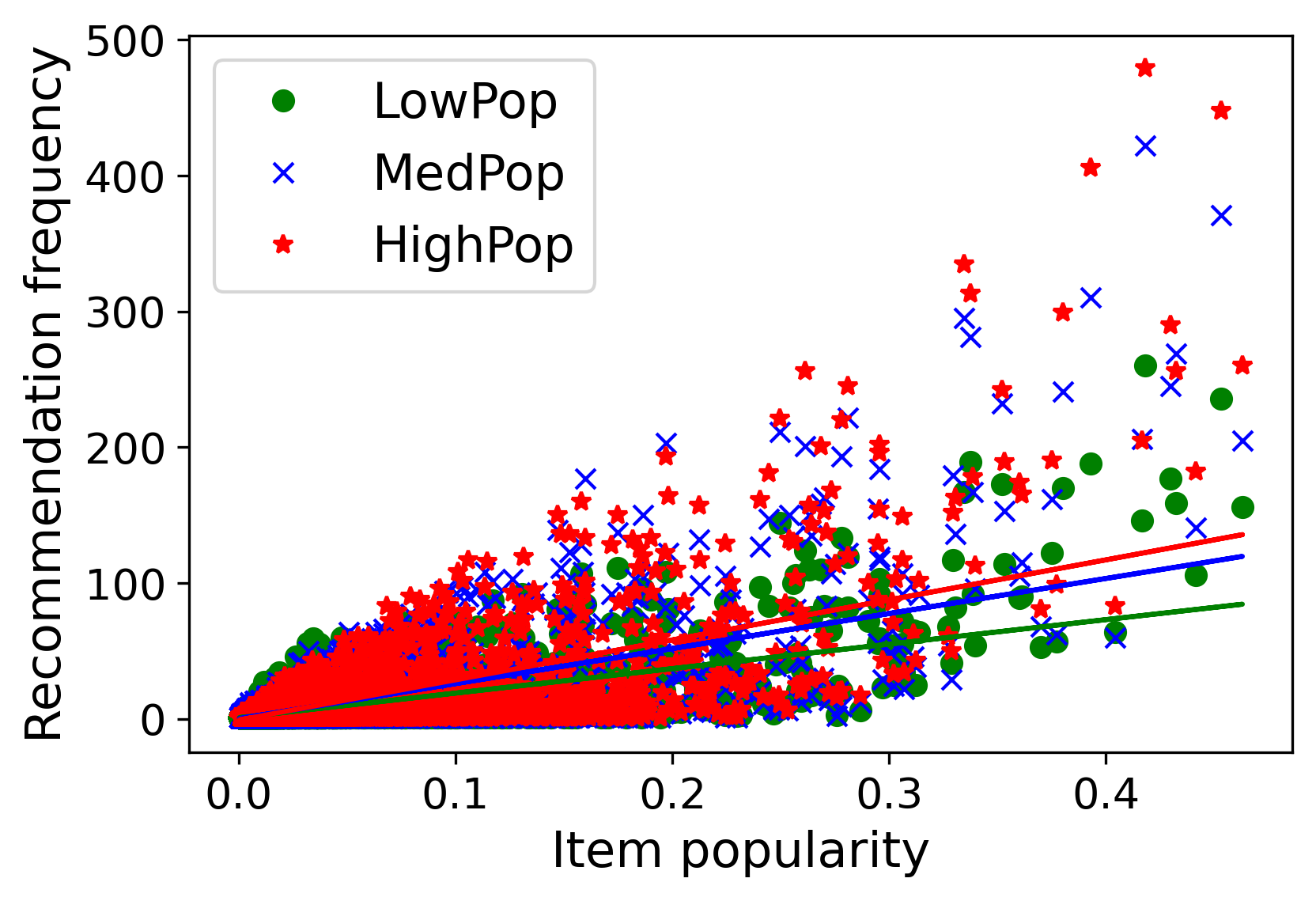}
& 
\includegraphics[width=.24\textwidth]{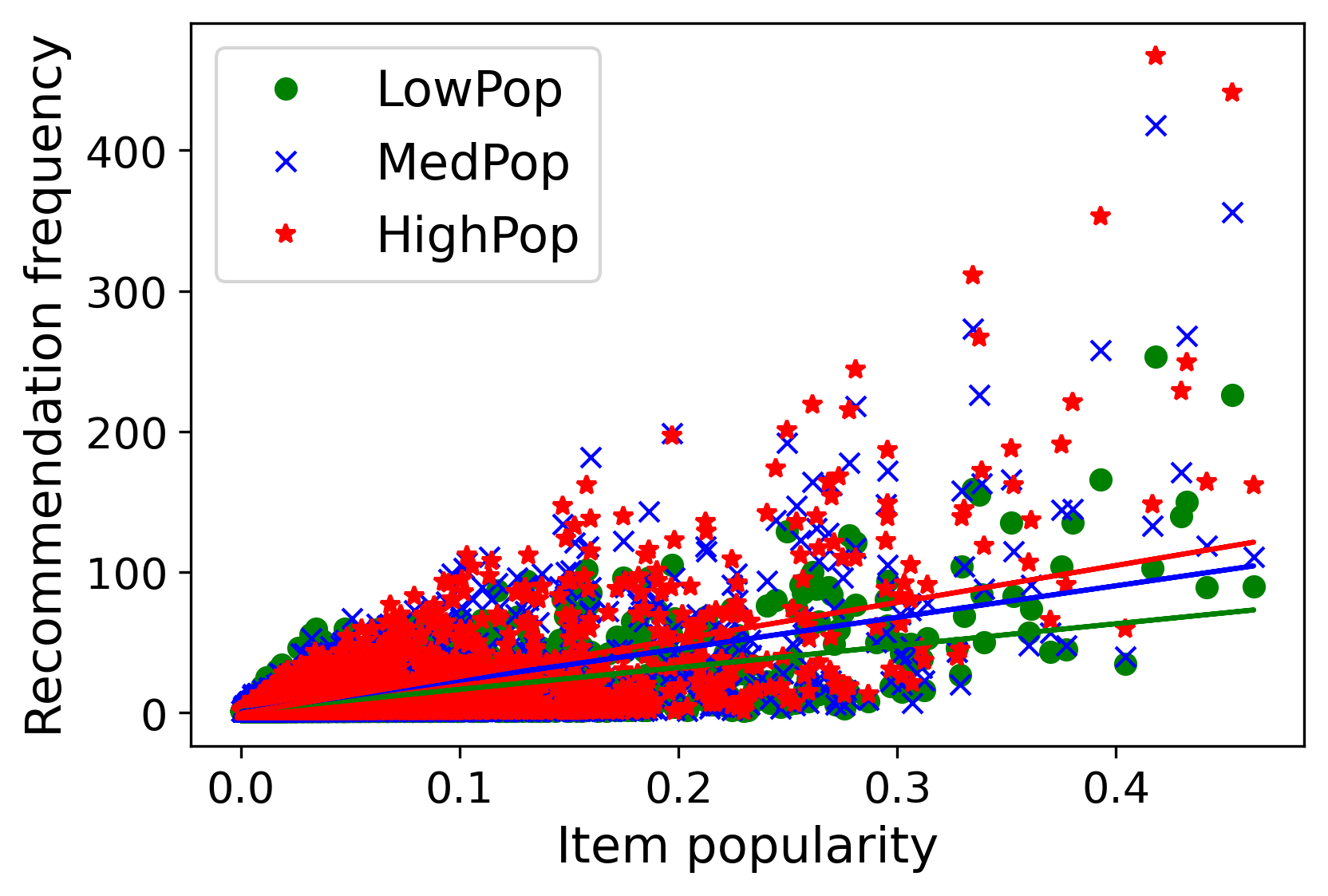}
&   
\includegraphics[width=.24\textwidth]{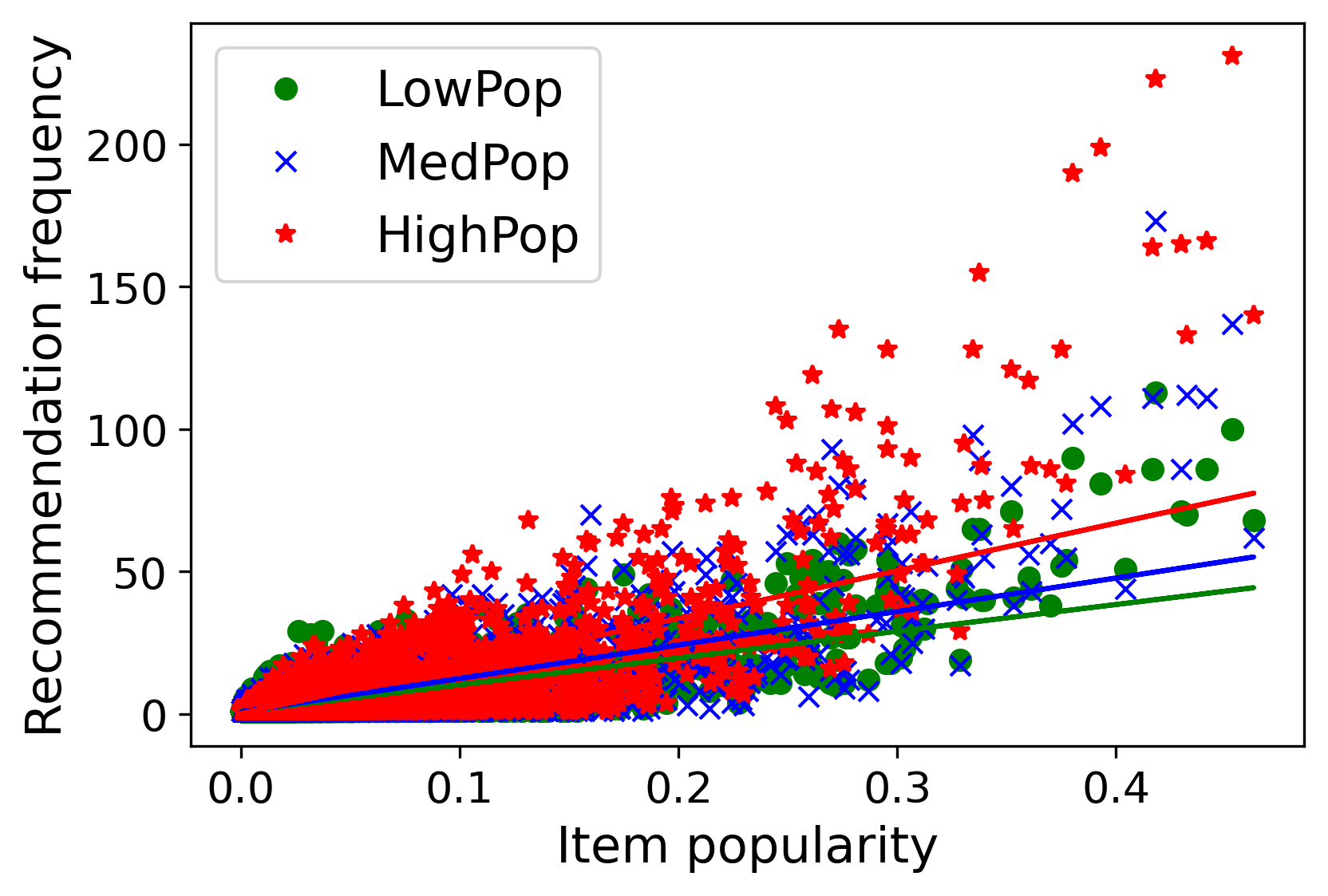}
&
\includegraphics[width=.24\textwidth]{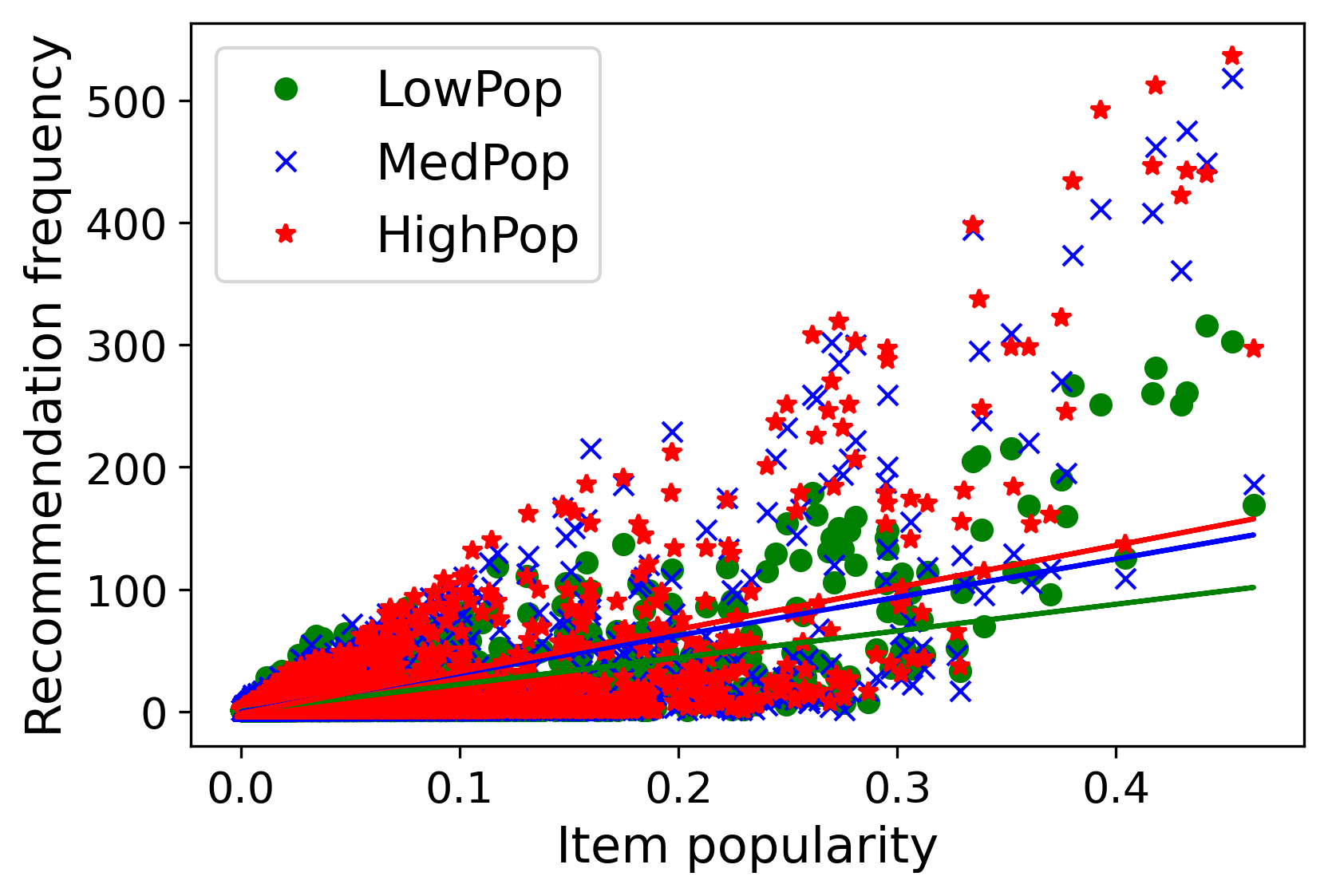}
&
\multirow{1}{*}[1.35cm]{\rotatebox{90}{Last.fm}}
\vspace{0.2cm}
\\ 
\includegraphics[width=.24\textwidth]{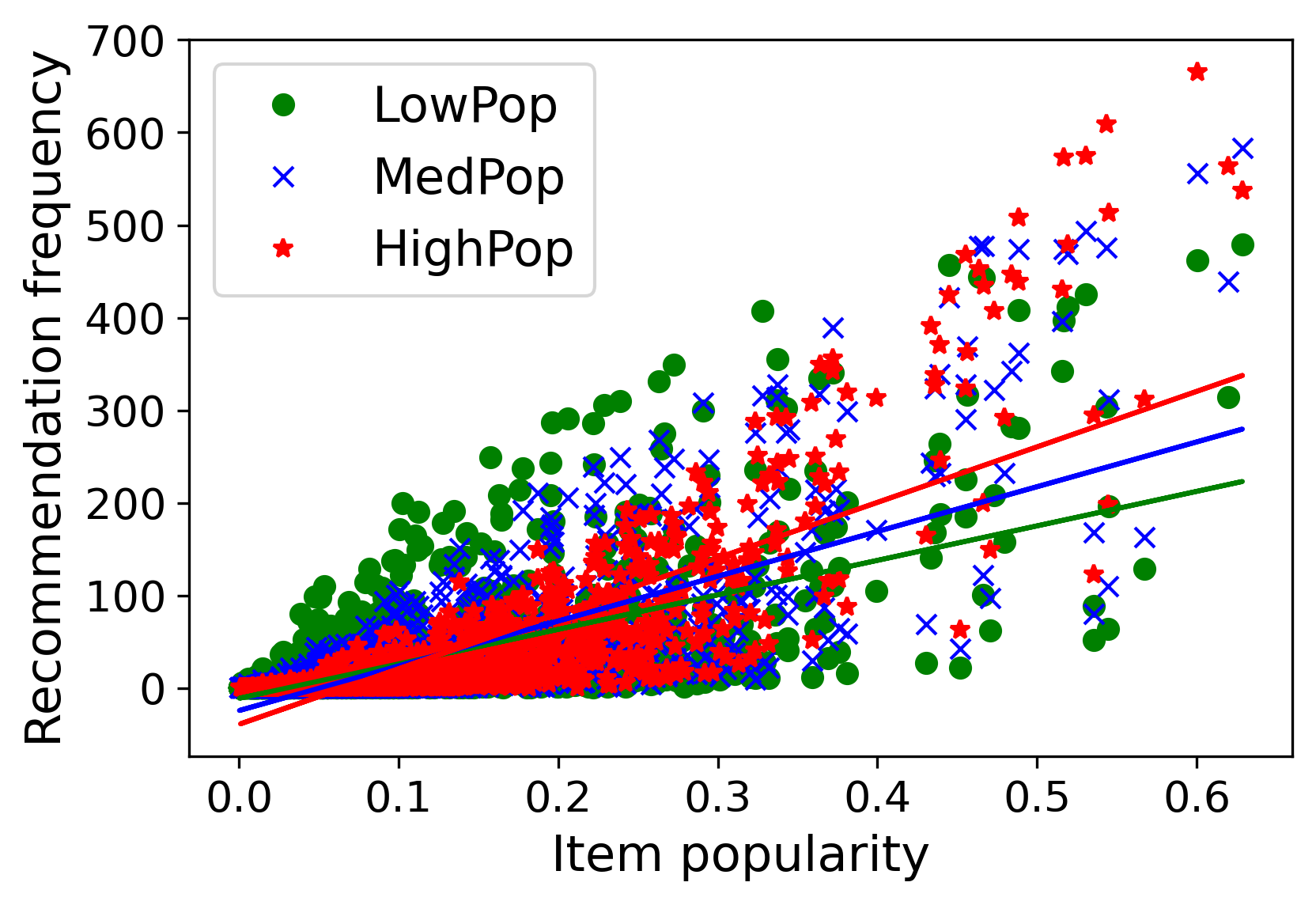}
& 
\includegraphics[width=.24\textwidth]{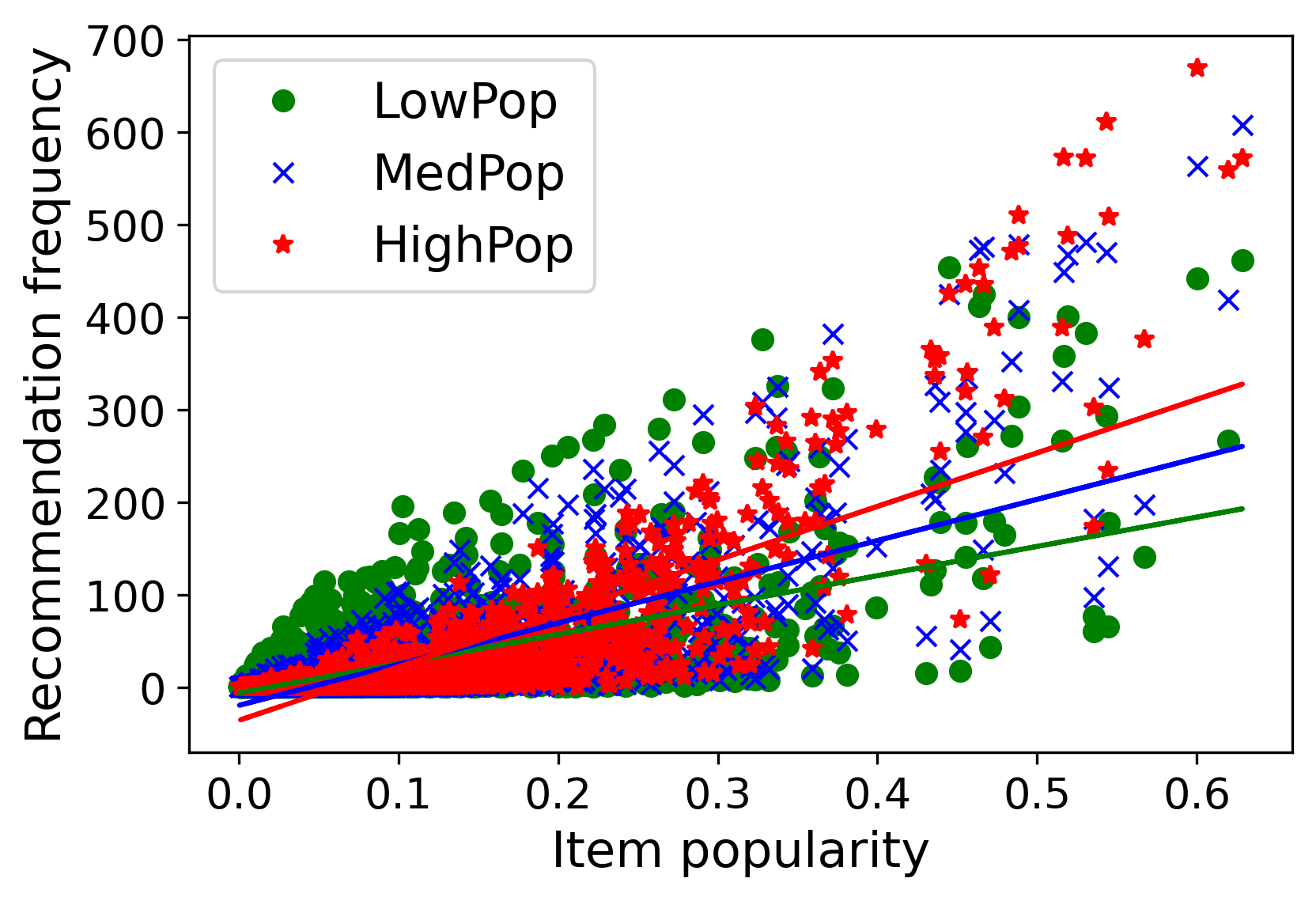}
&      
\includegraphics[width=.24\textwidth]{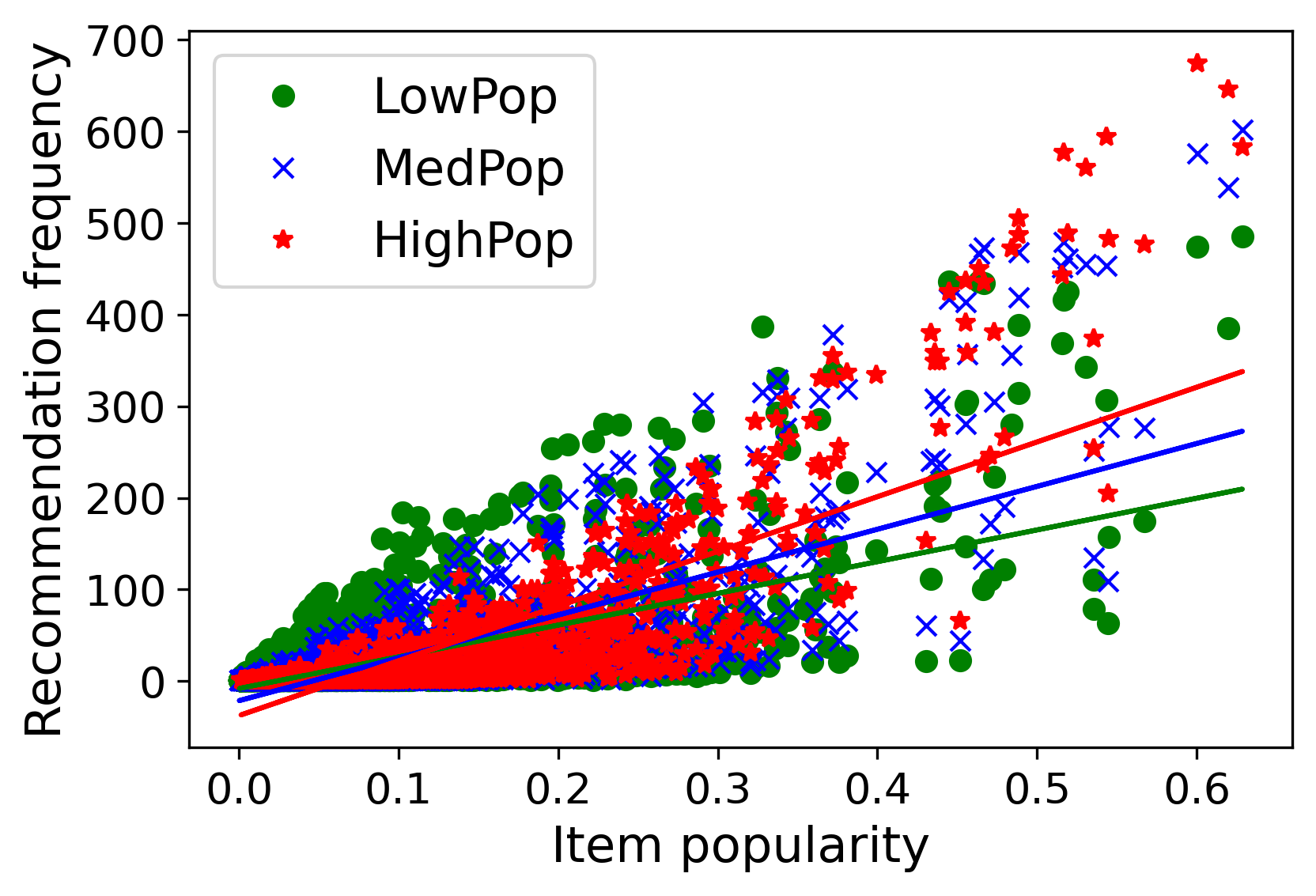}
&
\includegraphics[width=.24\textwidth]{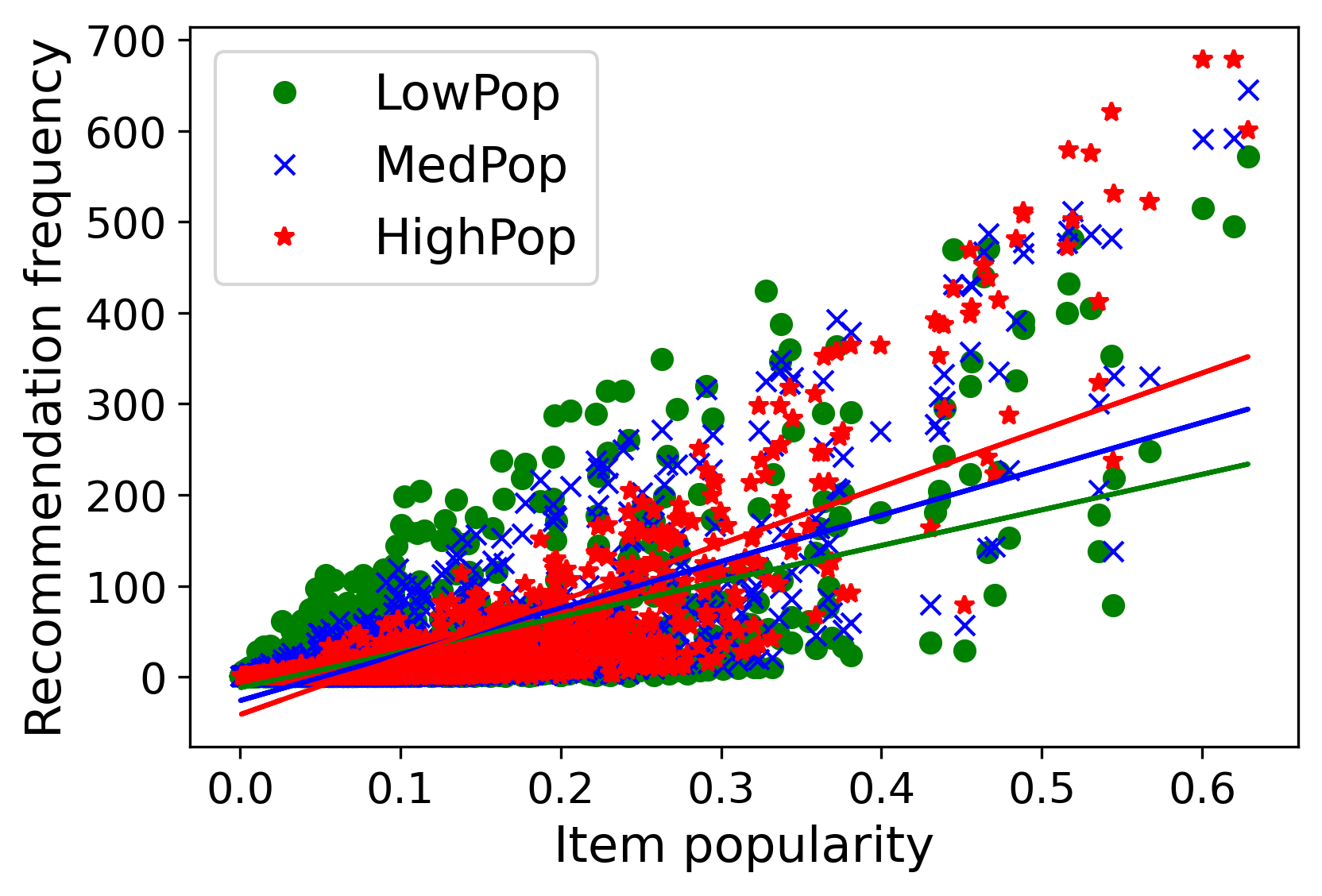}
&
\multirow{1}{*}[1.55cm]{\rotatebox{90}{MovieLens}}
\vspace{0.2cm}
\\ 
\includegraphics[width=.24\textwidth]{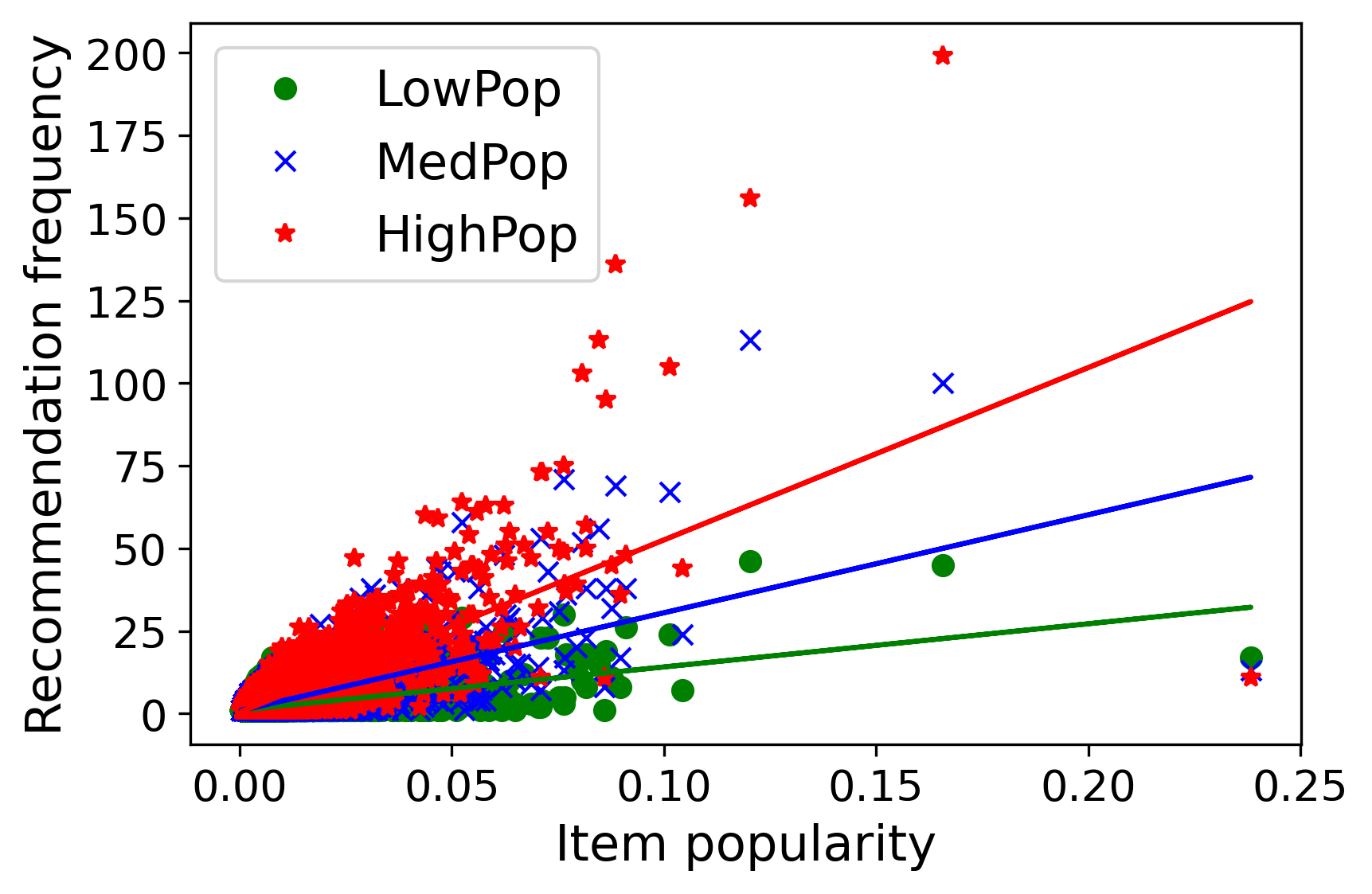}
& 
\includegraphics[width=.24\textwidth]{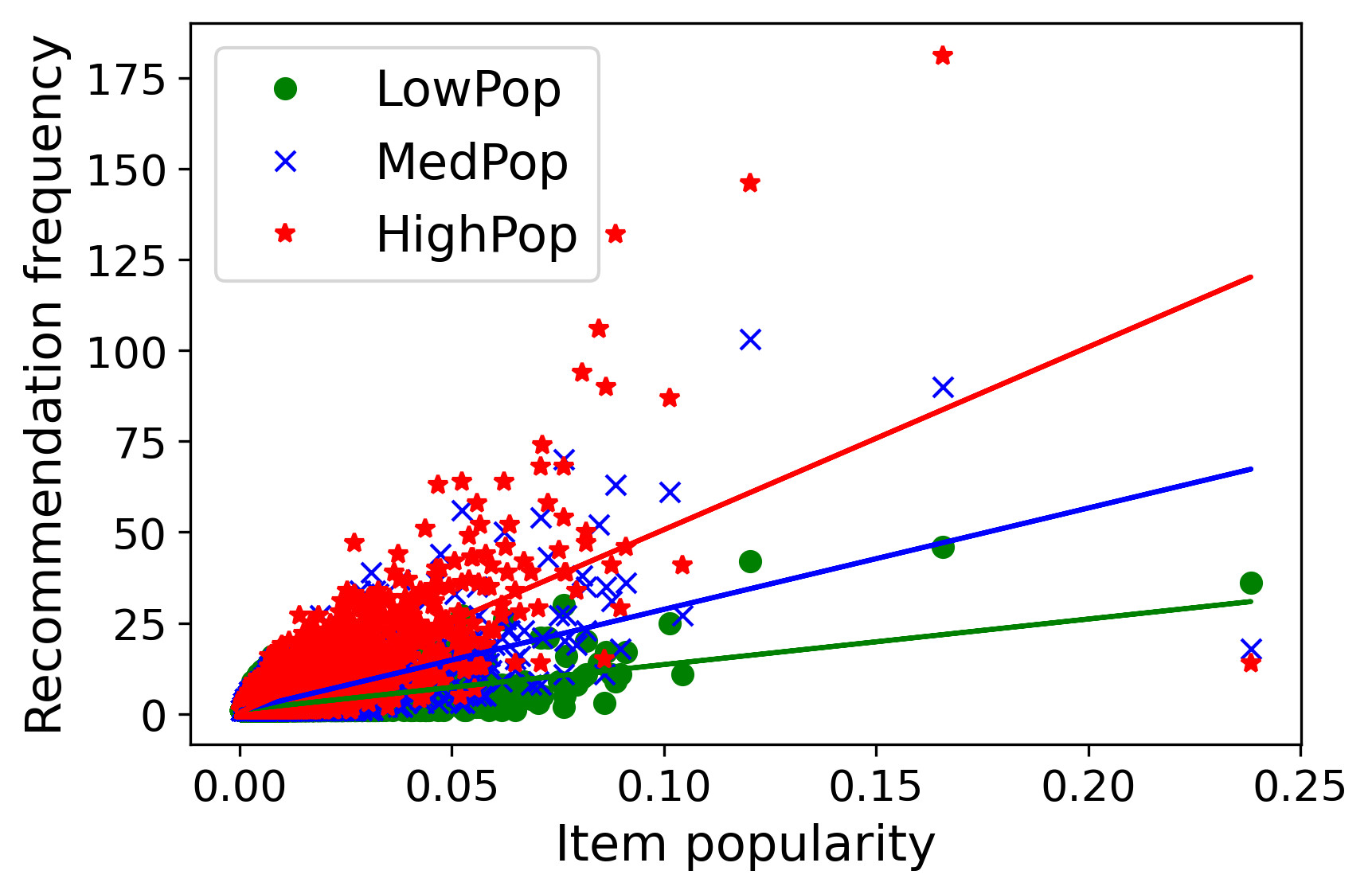}
&      
\includegraphics[width=.24\textwidth]{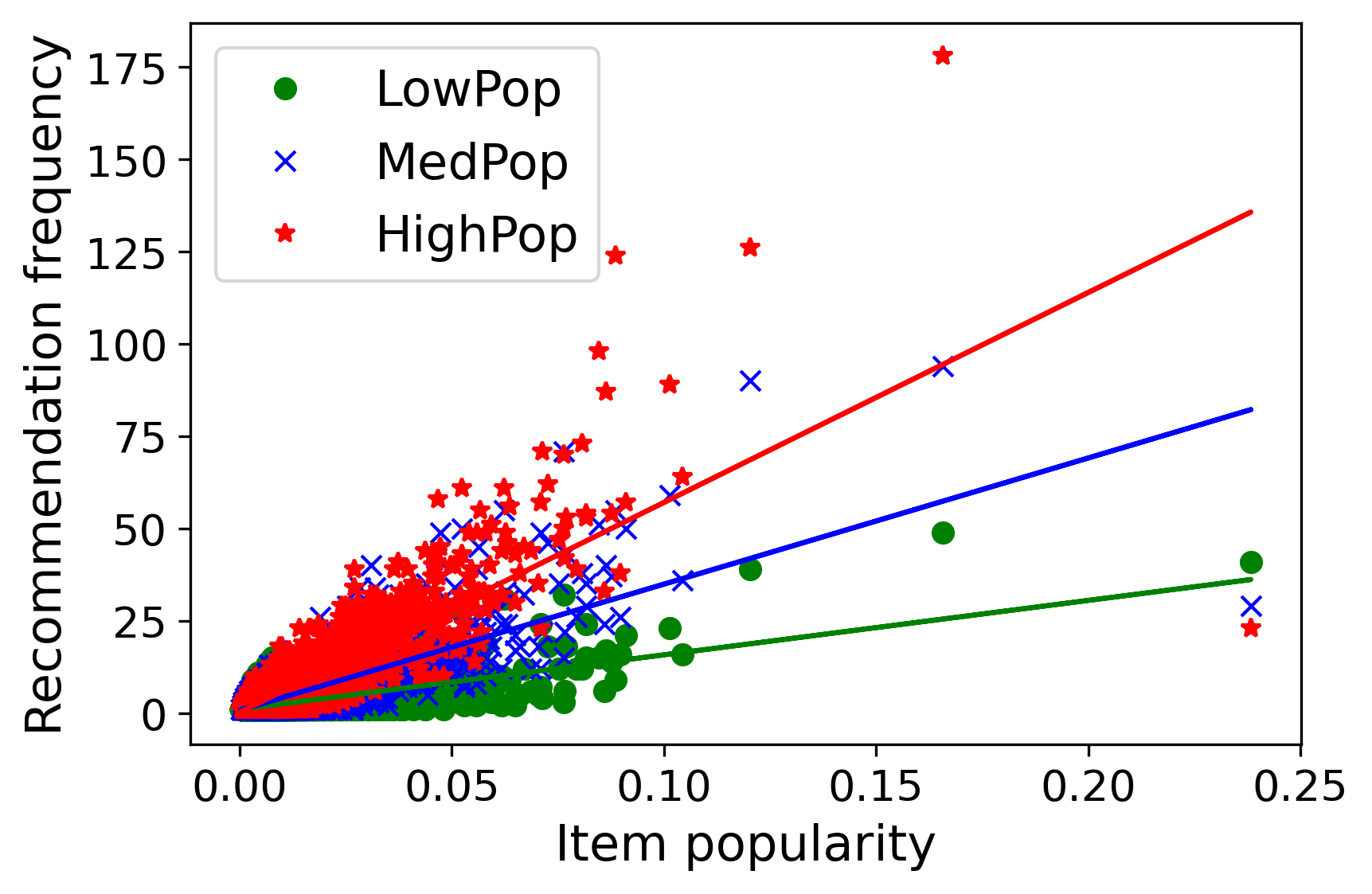}
&
\includegraphics[width=.24\textwidth]{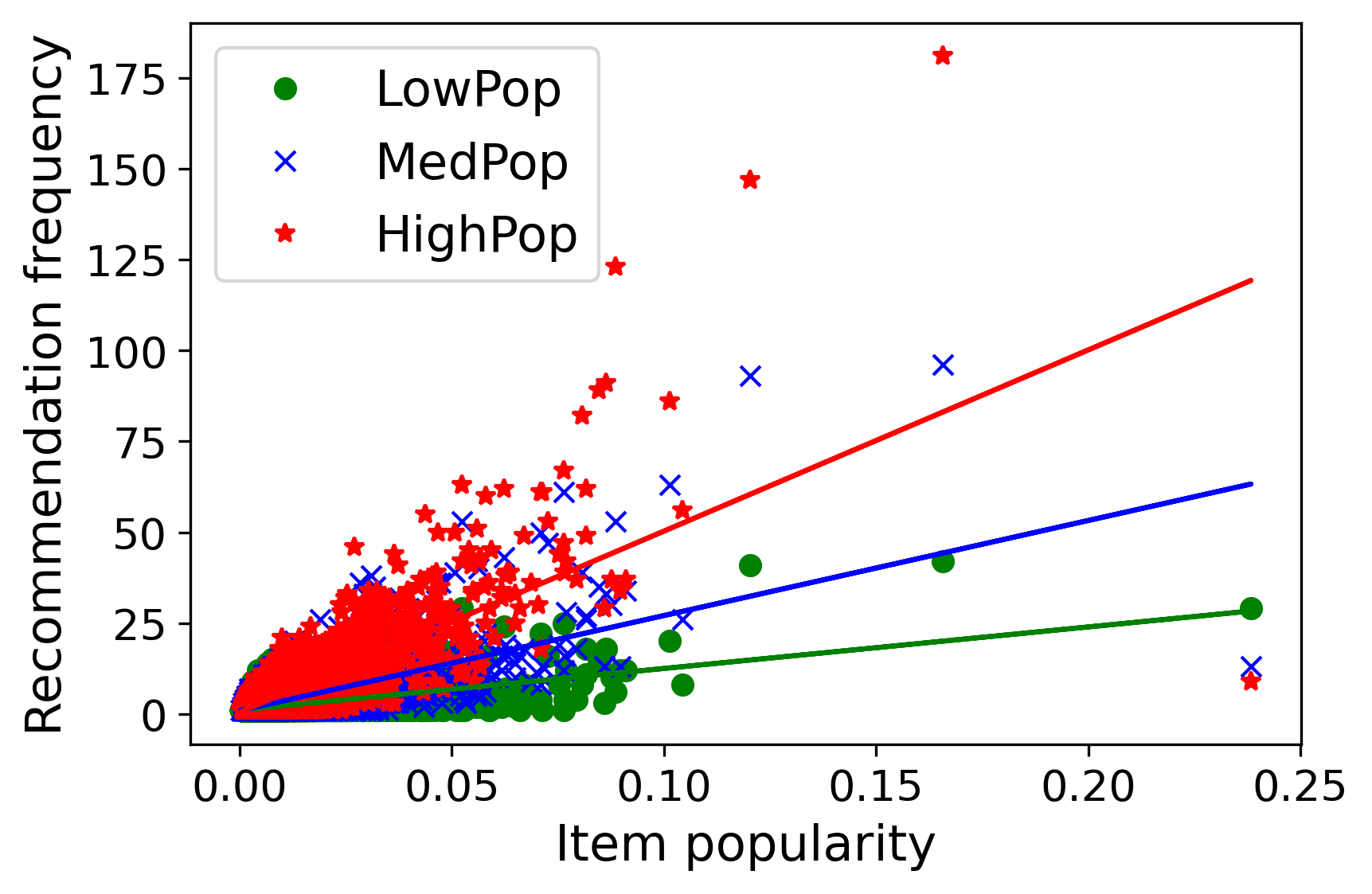}
&
\multirow{1}{*}[1.7cm]{\rotatebox{90}{BookCrossing}}
\vspace{0.2cm}
\\ 
\includegraphics[width=.24\textwidth]{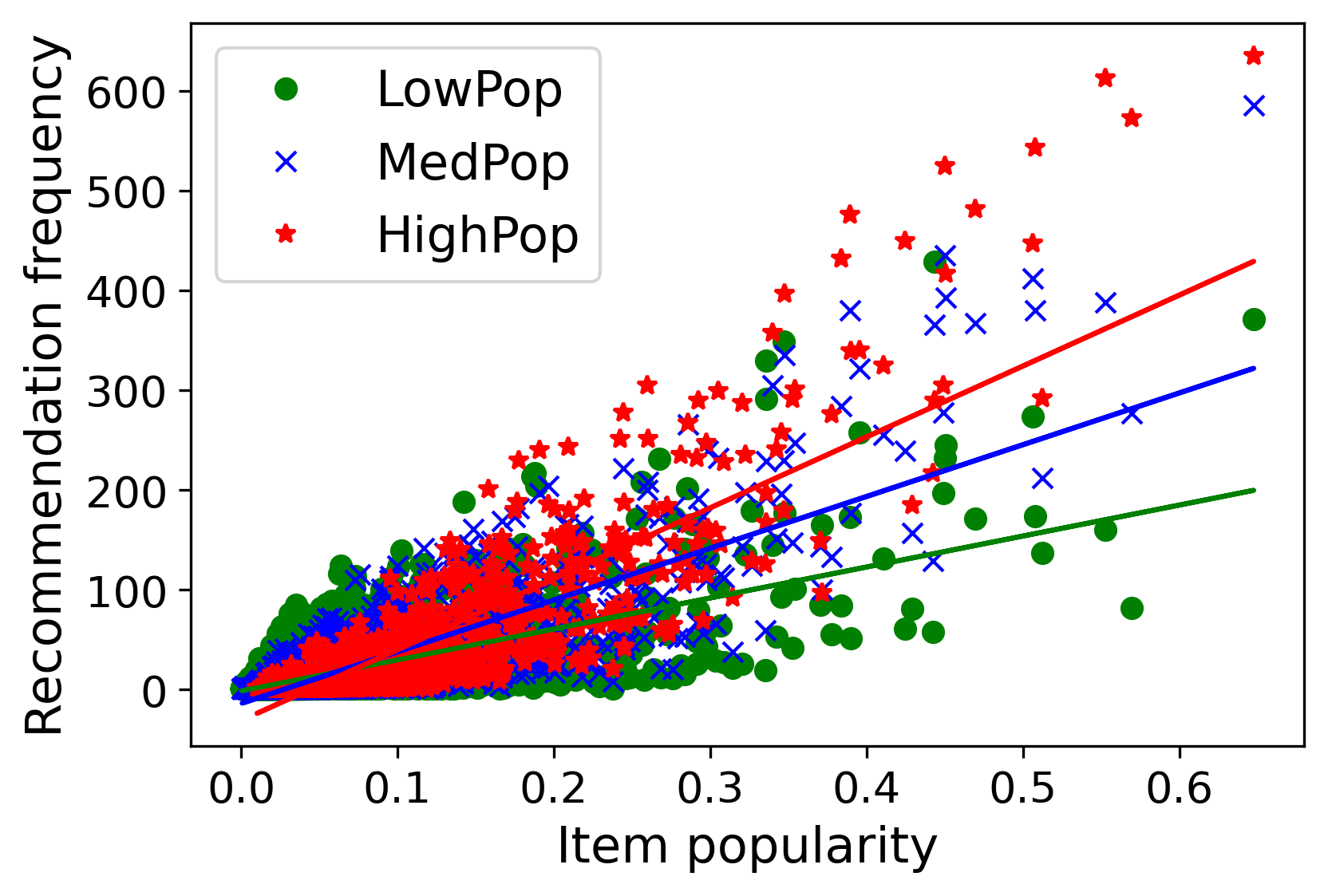}
& 
\includegraphics[width=.24\textwidth]{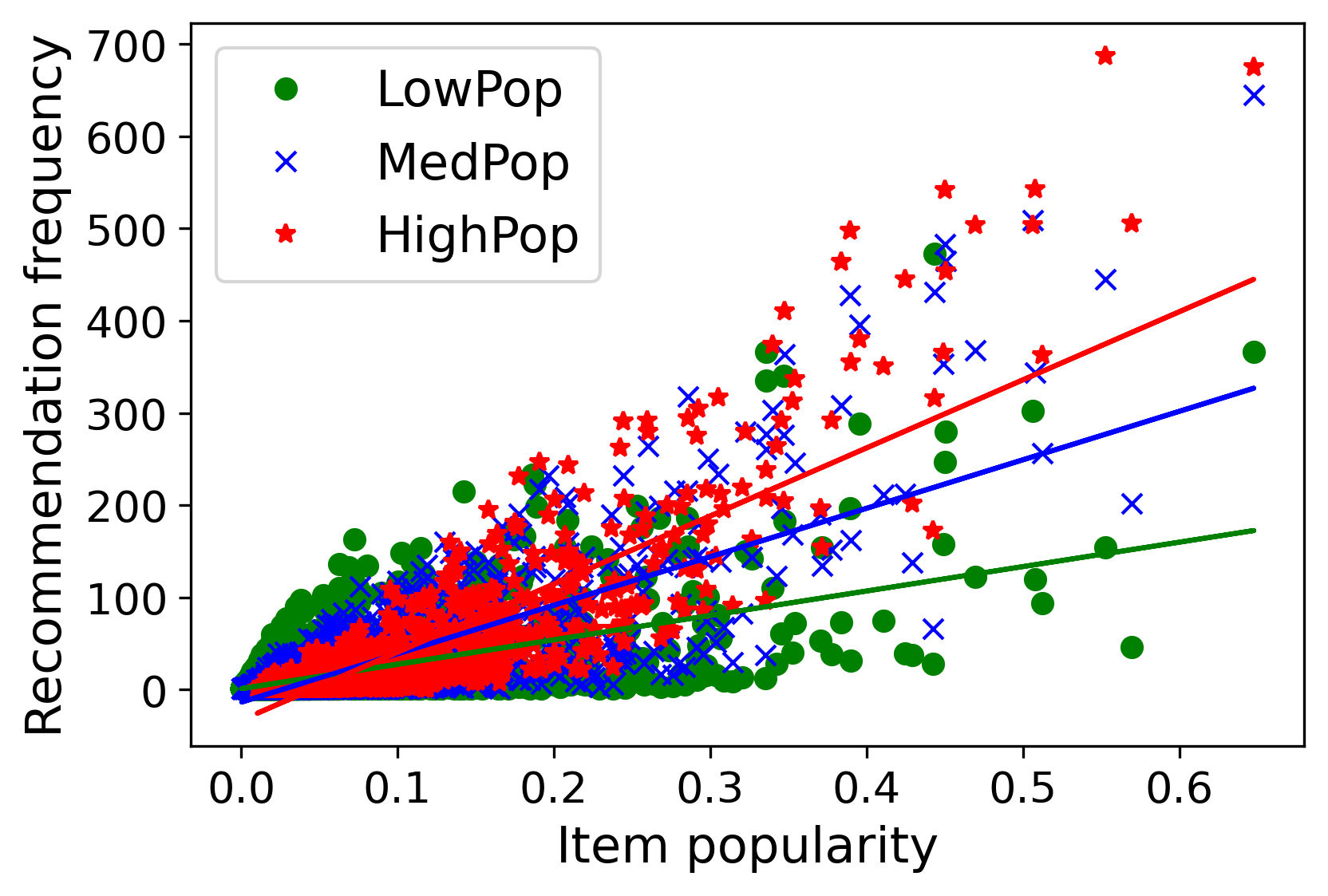}
&      
\includegraphics[width=.24\textwidth]{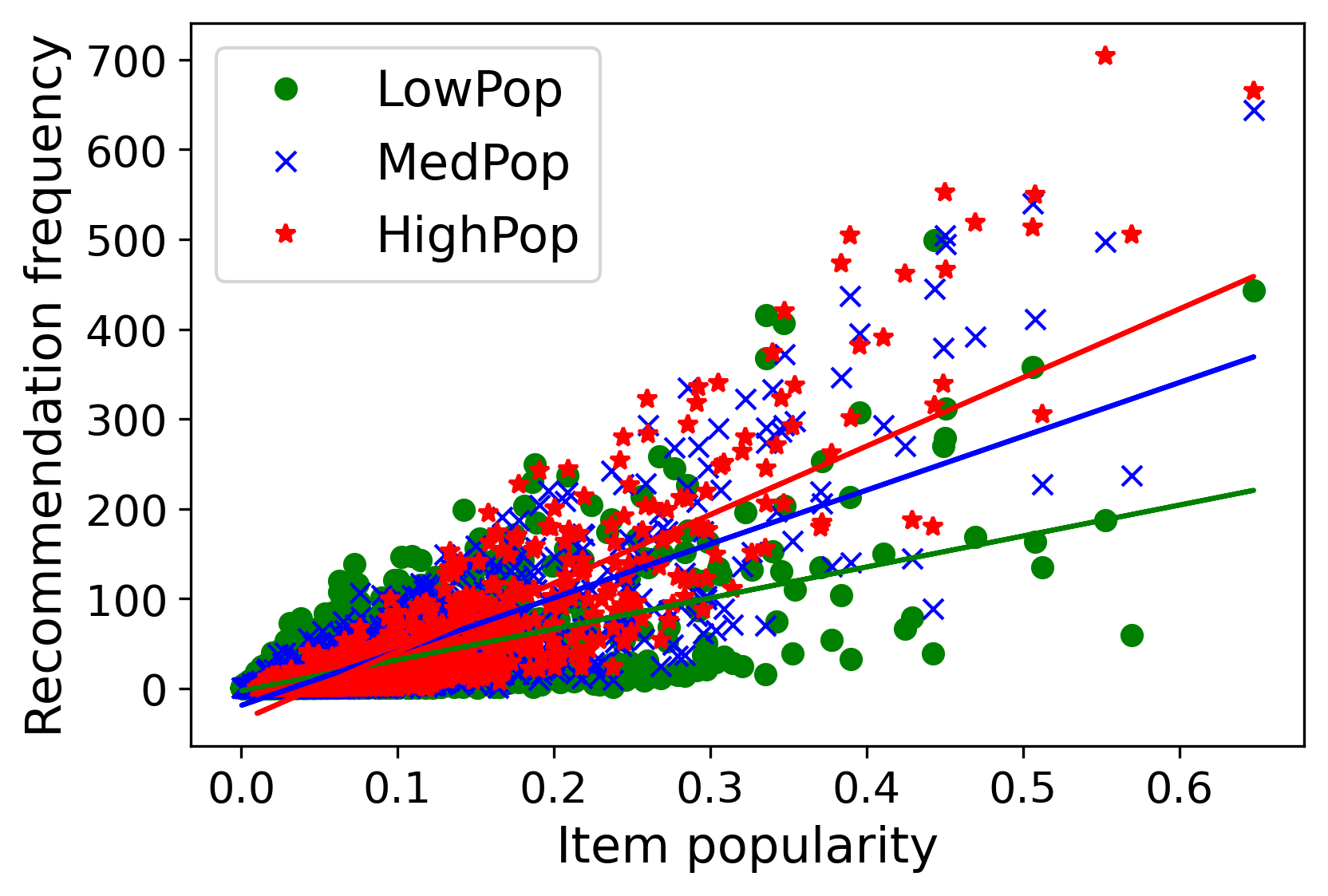}
&
\includegraphics[width=.24\textwidth]{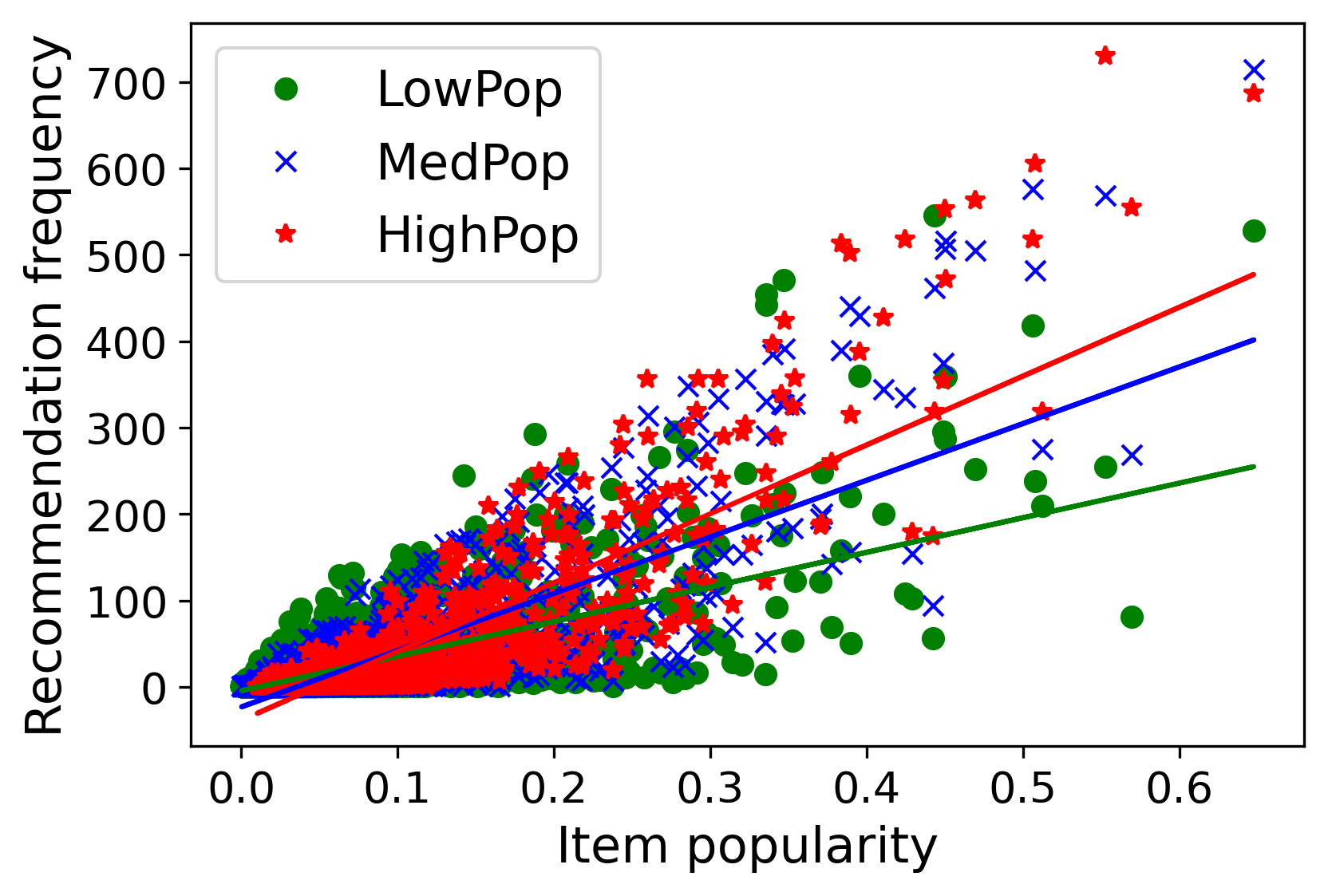}
&
\multirow{1}{*}[1.7cm]{\rotatebox{90}{MyAnimeList}}
\\ 
\end{tabular}
\end{center}

\caption{\textbf{RQ1:} Relationship between item popularity and recommendation frequency of four CF-based algorithms for Last.fm, MovieLens, BookCrossing and MyAnimeList. In all 16 cases, we see that popular media items have a higher probability of being recommended than unpopular ones.\vspace{-3mm}}
\label{fig:corr}
\end{figure}

\subsection{RQ1: Relationship Between Item Popularity and Recommendation Frequency}

Figure~\ref{fig:corr} shows the relationship between item popularity and recommendation frequency for the four CF-based algorithms UserKNN,  UserKNNAvg, NMF and CoClustering on all five folds of our four multimedia datasets Last.fm, MovieLens, BookCrossing and MyAnimeList. The solid lines indicate the linear regression between the two variables for the three user groups. 

In all 16 plots, and all three user groups, we observe a positive relationship between an item's popularity and how often this item gets recommended (\textbf{RQ1}).  
However, for NMF applied to Last.fm, the maximum recommendation frequency is much lower as in case of the other algorithms. 
Thus, only in case of NMF applied to Last.fm, we see a weak relationship between popularity and recommendation frequency, while in all other cases, we see a strong relationship between these variables. 
This is in line with our previous related work investigating popularity bias in Last.fm~\cite{kowald2020unfairness}. 
When comparing the three user groups, we see the weakest relationship between the variables for LowPop and the strongest relationship for HighPop. We will refer to this finding when investigating \textbf{RQ2}. 

\subsection{RQ2: Relationship Between Users' Inclination to Popular Items and Recommendation Accuracy}

\begin{table}[t]
\setlength{\tabcolsep}{3.5pt}	
\centering
\caption{\textbf{RQ2:} Mean absolute error (MAE) results (the lower, the better) of our study. The lowest accuracy is always given for the LowPop user group (statistically significant according to a t-test with $p < 0.001$ as indicated by $^{***}$ and $p < 0.05$ as indicated by $^{**}$). Across the algorithms, the best results are indicated by \textbf{bold numbers} and across the user groups, the best results are indicated by \textit{italic numbers}.\vspace{-3mm}}  
\begin{tabular}{l|l|llll}
\specialrule{.2em}{.1em}{.1em}
Dataset & User group & UserKNN    & UserKNNAvg    & NMF  & CoClustering    \\\hline
\multirow{3}{*}{Last.fm} & LowPop    & 49.489$^{***}$    & 46.483$^{***}$  & \textbf{39.641$^{**}$}  & 47.304$^{***}$ \\
& MedPop                         & \textit{42.899}   & \textit{37.940}   & \textit{\textbf{32.405}}        & \textit{37.918} \\
& HighPop                        & 45.805            & 43.070        & \textbf{38.580}        & 42.982 \\\hline

\multirow{3}{*}{MovieLens} & LowPop     & 0.801$^{***}$     & 0.763$^{***}$  & 0.753$^{***}$   & \textbf{0.738$^{***}$} \\
& MedPop                         & 0.748     & 0.727        & 0.722        & \textbf{0.705} \\
& HighPop                        & \textit{0.716}     & \textit{0.697}        & \textit{0.701}   & \textit{\textbf{0.683}} \\\hline

\multirow{3}{*}{BookCrossing} & LowPop     & 1.403$^{***}$     & \textbf{1.372$^{***}$}        & 1.424$^{***}$  & 1.392$^{***}$ \\
& MedPop                         & \textit{1.154}     & \textit{\textbf{1.122}}        & \textit{1.214}        & \textit{1.134} \\
& HighPop                        & 1.206             & \textbf{1.155}                & 1.274         & 1.162 \\\hline

\multirow{3}{*}{MyAnimeList} & LowPop    & 1.373$^{***}$     & \textbf{1.001$^{***}$}    & 1.010$^{***}$     & 1.001$^{***}$ \\
& MedPop                         & 1.341     &        \textbf{0.952}             & 0.968             & \textit{0.956} \\
& HighPop                        & \textit{1.311}     & \textit{\textbf{0.948}}  & \textit{0.951}    & 0.975 \\
\specialrule{.2em}{.1em}{.1em}
\end{tabular}
\label{tab:mae}
\end{table}

Table~\ref{tab:mae} shows the MAE estimates for the aforementioned CF-based recommendation algorithms (UserKNN, UserKNNAvg, NMF, and CoClustering) on the four multimedia datasets (Last.fm, MovieLens, BookCrossing, and MyAnimeList) split in three user groups that differ in their inclination to popularity (LowPop, MedPop, and HighPop). 
Additionally, we indicate statistically significant differences between both LowPop and MedPop, and LowPop and HighPop according to a t-test with $p < 0.001$ using $^{***}$ and with $p < 0.05$ using $^{**}$ in the LowPop lines. 

Across all datasets, we observe the highest MAE estimates, and thus lowest recommendation accuracy, for the LowPop user groups. The best results, indicated by \textit{italic numbers}, are reached for the MedPop group in case of Last.fm and BookCrossing, and for the HighPop group in case of MovieLens and MyAnimeList. For Last.fm this is in line with our previous work~\cite{kowald2020unfairness}. 
Across the algorithms, we see varying results: for Last.fm, and again in line with our previous work~\cite{kowald2020unfairness}, the best results are reached for NMF. For MovieLens, we get the best results for the CoClustering approach, and for BookCrossing and MyAnimeList the highest accuracy is reached for the UserKNN variant UserKNNAvg. We plan to investigate these differences across the user groups and the algorithms in our future research, as outlined in the next section.

Taken together, users with little inclination to popular multimedia items receive statistically significantly worse recommendations by CF-based algorithms than users with medium and high inclination to popularity (\textbf{RQ2}). 
When referring back to our results of \textbf{RQ1} in Figure~\ref{fig:corr}, this is interesting since LowPop is the group with the weakest relationship between item popularity and recommendation frequency. However, this suggests that recommendations are still too popular for this user group and an adequate mitigation strategy is needed.

\section{Conclusion}
\label{s:conc}

In this paper, we have studied popularity bias in CF-based MMRS. 
Therefore, we investigated four recommendation algorithms (UserKNN, UserKNNAvg, NMF, and CoClustering) for three user groups (LowPop, MedPop, and HighPop) on four multimedia datasets (Last.fm, MovieLens, BookCrossing, and MyAnimeList). 
Specifically, we investigated popularity bias from the item (\textbf{RQ1}) and user (\textbf{RQ2}) perspective.
Additionally, we have shown that users with little interest into popular items tend to have large profile sizes, and therefore are important data sources for MMRS.

With respect to \textbf{RQ1}, we find that the popularity of a multimedia item strongly correlates with the probability that this item is recommended by CF-based approaches. 
With respect to \textbf{RQ2}, we find that users with little interest in popular multimedia items (i.e., LowPop) receive significantly worse recommendations than users with medium (i.e., MedPop) or high (i.e., HighPop) interest in popular items. This is especially problematic since users with little interest into popularity tend to have large profile sizes, and thus, should be treated in a fair way by MMRS.

\para{Future work.} Our results demonstrate that future work should further focus on studying this underserved user group in order to mitigate popularity bias in CF-based recommendation algorithms. We believe that our findings are a first step to inform the research on popularity bias mitigation techniques (see Section~\ref{s:relwork}) to choose the right mitigation strategy for a given setting.

Additionally, as mentioned earlier, we plan to further study the differences we found with respect to algorithmic performance for the different user groups and multimedia domains. Here, we also want to study popularity bias in top-$n$ settings using ranking-aware metrics such as nDCG (e.g., as used in~\cite{lacic2015tackling}).
Finally, we plan to work on further bias mitigation strategies based on cognitive-inspired user modeling and recommendation techniques (e.g.,~\cite{www_sustain,lacic2014recommending,Kowald2016bllhypertext}.

\para{Acknowledgements.} This research was funded by the H2020 project TRUSTS (GA: 871481)  and the ``DDAI'' COMET Module within the COMET – Competence Centers for Excellent Technologies Programme, funded by the Austrian Federal Ministry for Transport, Innovation and Technology (bmvit), the Austrian Federal Ministry for Digital and Economic Affairs (bmdw), the Austrian Research Promotion Agency (FFG), the province of Styria (SFG) and partners from industry and academia. The COMET Programme is managed by FFG.

\bibliographystyle{splncs04}
\bibliography{bib}

\end{document}